%% file: main.tex
\documentclass[sigconf]{acmart}
\AtBeginDocument{%
  }

\copyrightyear{2026}
\acmYear{2026}
\setcopyright{cc}
\setcctype{by}
\acmConference[CHI '26]{Proceedings of the 2026 CHI Conference on Human Factors in Computing Systems}{April 13--17, 2026}{Barcelona, Spain}
\acmBooktitle{Proceedings of the 2026 CHI Conference on Human Factors in Computing Systems (CHI '26), April 13--17, 2026, Barcelona, Spain}
\acmDOI{10.1145/3772318.3791966}
\acmISBN{979-8-4007-2278-3/2026/04}

\usepackage[nameinlink]{cleveref}
\usepackage{tikz}
\usepackage{xspace}
\usepackage{acmart-taps}
\usepackage{graphicx}

\aptLtoXcmd{\newcommand{\textlabel}[3]{}}%
{\newcommand{\textlabel}[3][Black]{%
  \tikz[baseline=(X.base)]
    \node[rectangle, rounded corners=2pt, fill=#1!25, text=#2, inner xsep=1.5pt, inner ysep=1.5pt] (X) {\texttt {#3}};\xspace%
}}

\definecolor{ProdBlue}{HTML}{5F90D3}
\definecolor{DSysPurple}{HTML}{A164DF}
\definecolor{FeatOrange}{HTML}{F79E5D}
\definecolor{CompTeal}{HTML}{5BD1B7}

\definecolor{ProdBlueText}{HTML}{1C398E}
\definecolor{DSysPurpleText}{HTML}{59168B}
\definecolor{FeatOrangeText}{HTML}{9F2D00}
\definecolor{CompTealText}{HTML}{0D4E4A}

\aptLtoXcmd{\long\def\ProductTag#1{{\xbox{aptbox}{\XMLaddatt{style}{background-color:\#e4ecf8; color:\#1C398E;border-radius:5px;padding:4px;white-space:nowrap;}\textsand{#1}}}\xspace}}{\newcommand{\ProductTag}[1]{{\textlabel[ProdBlue]{ProdBlueText}{#1}}\xspace}}

\aptLtoXcmd{\long\def\DesignSystemTag#1{{\xbox{aptbox}{\XMLaddatt{style}{background-color:\#efe6fb;color:\#59168B;border-radius:5px;padding:4px;white-space:nowrap;}\textsand{#1}}}\xspace}}{\newcommand{\DesignSystemTag}[1]{{\textlabel[DSysPurple]{DSysPurpleText}{#1}}\xspace}}

\aptLtoXcmd{\long\def\FeatureTag#1{{\xbox{aptbox}{\XMLaddatt{style}{background-color:\#fff1e6;color:\#9F2D00;border-radius:5px;padding:4px;white-space:nowrap;}\textsand{#1}}}\xspace}}{\newcommand{\FeatureTag}[1]{{\textlabel[FeatOrange]{FeatOrangeText}{#1}}\xspace}}

\aptLtoXcmd{\long\def\ComponentTag#1{{\xbox{aptbox}{\XMLaddatt{style}{background-color:\#e6f7f3;color:\#0D4E4A;border-radius:5px;padding:4px;white-space:nowrap;}\textsand{#1}}}\xspace}}{\newcommand{\ComponentTag}[1]{{\textlabel[CompTeal]{CompTealText}{#1}}\xspace}}

\newcommand{\eg}{\textit{e.g.}\xspace}
\newcommand{\etal}{\textit{et al.}\xspace}

\begin{document}

\title{Bridging Gulfs in UI Generation through Semantic Guidance}

\author{Seokhyeon Park}
\orcid{0009-0003-1685-4027}
\affiliation{%
  \institution{Seoul National University}
  \city{Seoul}
  \country{Republic of Korea}
}
\email{shpark@hcil.snu.ac.kr}

\author{Soohyun Lee}
\orcid{0000-0002-3075-3981}
\affiliation{%
  \institution{Seoul National University}
  \city{Seoul}
  \country{Republic of Korea}
}
\email{shlee@hcil.snu.ac.kr}

\author{Eugene Choi}
\orcid{0009-0001-5189-6367}
\affiliation{%
  \institution{Seoul National University}
  \city{Seoul}
  \country{Republic of Korea}
}
\email{eugene@snu.ac.kr}

\author{Hyunwoo Kim}
\orcid{0009-0007-1827-165X}
\affiliation{%
  \institution{Seoul National University}
  \city{Seoul}
  \country{Republic of Korea}
}
\email{hyunwoo0628@snu.ac.kr}

\author{Minkyu Kweon}
\orcid{0009-0000-2557-6055}
\affiliation{%
  \institution{Seoul National University}
  \city{Seoul}
  \country{Republic of Korea}
}
\email{mk@hcil.snu.ac.kr}

\author{Yumin Song}
\orcid{0009-0004-5277-4822}
\affiliation{%
  \institution{Seoul National University}
  \city{Seoul}
  \country{Republic of Korea}
}
\email{ymsong@hcil.snu.ac.kr}

\author{Jinwook Seo}
\orcid{0000-0002-7734-822X}
\authornote{Corresponding Author}
\affiliation{%
  \institution{Seoul National University}
  \city{Seoul}
  \country{Republic of Korea}
}
\email{jseo@snu.ac.kr}

\renewcommand{\shortauthors}{Park et al.}

\begin{abstract}
\input{section/0-abstract.tex}
\end{abstract}


\begin{teaserfigure}
  \includegraphics[width=\textwidth]{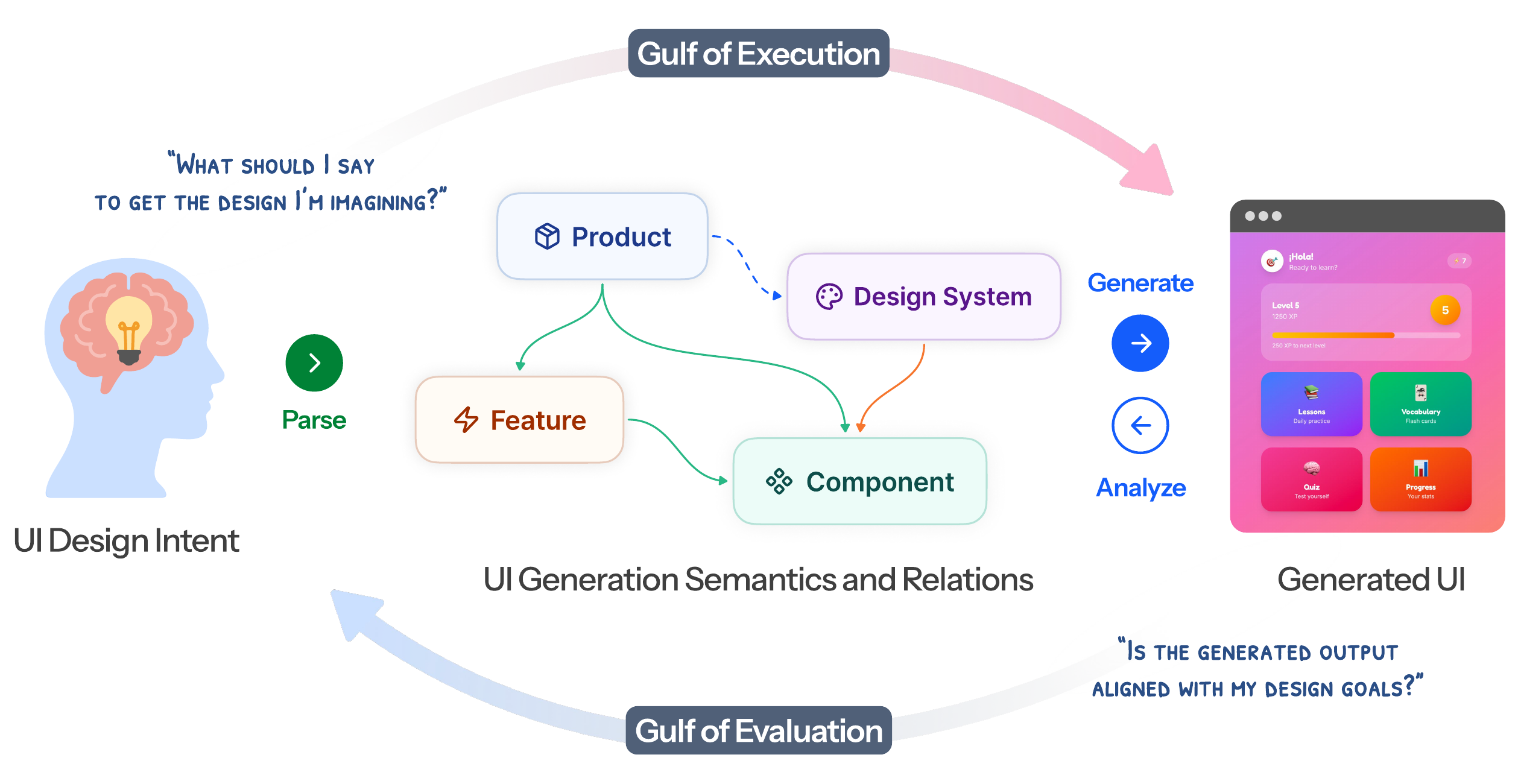}
  \caption{\textit{Bridging gulfs of execution and evaluation in generative UI design.} \textmd{While text-to-UI systems enable high-fidelity interface generation, users face challenges in articulating design intent and interpreting results. Our approach introduces explicit semantic representations as an intermediate layer between human intent and AI output. By \protect\raisebox{-0.15em}{\includegraphics[height=0.9em]{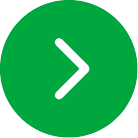}} parsing input semantics, \protect\raisebox{-0.15em}{\includegraphics[height=0.9em]{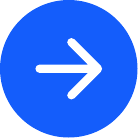}} generating interfaces, and \protect\raisebox{-0.15em}{\includegraphics[height=0.9em]{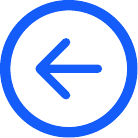}} analyzing output semantics and their relations, our system makes requirements explicit and outcomes interpretable, supporting more systematic, explainable, and iterative UI generation.}}
  \Description{System overview showing how user design intent is parsed into semantic categories, which generate a UI and support analysis. The diagram highlights bridging the Gulf of Execution and Evaluation by making design intent explicit and outputs interpretable.}
  \label{fig:teaser}
\end{teaserfigure}

\maketitle

\input{section/1-introduction}
\input{section/2-relatedwork}

\input{section/3-gulfs}
\input{section/4-semantics}
\input{section/5-system}
\input{section/6-evaluation}

\input{section/7-discussion}
\input{section/8-conclusion}

\begin{acks}
This work was supported by the National Research Foundation of Korea (NRF) grant funded by the Korean government (MSIT) (No. 2023R1A2C200520911) and the Institute of Information \& Communications Technology Planning \& Evaluation (IITP) grant funded by the Korean government (MSIT) [No. RS-2021-II211343, Artificial Intelligence Graduate School Program (Seoul National University)], and by the SNU-Global Excellence Research Center establishment project. The ICT at Seoul National University provided research facilities for this study.
\end{acks}

\bibliographystyle{ACM-Reference-Format}
\bibliography{references}

\end{document}

%% file: section/0-abstract.tex
While generative AI enables high-fidelity UI generation from text prompts, users struggle to articulate design intent and evaluate or refine results—creating gulfs of execution and evaluation.
To understand the information needed for UI generation, we conducted a thematic analysis of UI prompting guidelines, identifying key design semantics and discovering that they are hierarchical and interdependent.
Leveraging these findings, we developed a system that enables users to specify semantics, visualize relationships, and extract how semantics are reflected in generated UIs.
By making semantics serve as an intermediate representation between human intent and AI output, our system bridges both gulfs by making requirements explicit and outcomes interpretable.
A comparative user study suggests that our approach enhances users' perceived control over intent expression and outcome interpretation, and facilitates more predictable iterative refinement. 
Our work demonstrates how explicit semantic representation enables systematic and explainable exploration of design possibilities in AI-driven UI design.

%% file: section/1-introduction.tex
\section{Introduction}

The rapid advancement of generative AI has introduced new possibilities in the landscape of user interface (UI) design~\cite{nielsen-paradigm}.
Today, users can generate high-fidelity user interfaces through natural language prompts, democratizing access to sophisticated design capabilities~\cite{genuistudy}.
However, this apparent simplicity masks significant challenges in bidirectional communication between users and AI systems: users struggle to effectively express their design intentions, interpret generated outputs, and predict the effects of modifications.

These challenges manifest in three interconnected ways that limit the technology's potential. 
First, the complexity of UI design creates a multifaceted space of considerations, leaving users struggling to determine what information to specify and how to represent it in their prompts.
User interfaces are inherently multifaceted artifacts where visual aesthetics, interaction patterns, information architecture, and user context must harmoniously converge~\cite{garrett2010elements, s&ui, yuwen-bridging}, and these elements are interdependent, meaning changes to one often affect others.
Current AI systems provide no guidance on this information landscape, leaving users uncertain about what to include.
Furthermore, even when users know what they want, translating design intent into text remains problematic.
For example, users often use conceptual descriptors~\cite{designingwithai} such as ``modern,'' which can mean flat design, material design, glassmorphism, or numerous other design languages, each with distinct visual characteristics.
This dual challenge—not knowing what to specify and lacking the vocabulary to express it—results in underspecified or ambiguous prompts that fail to capture design requirements.
Second, users face an opaque generation process where countless design decisions have been made without explanation.
Without visibility into how inputs translated to design outputs, users cannot effectively evaluate whether results align with their intentions---they have to reverse-engineer the AI's interpretation rather than directly assess designs.
Third, these challenges compound through iteration, a core principle of UI design.
Each refinement cycle begins with an imperfect understanding of the output, leading to imprecise modification requests, which produce even less predictable results. 
Also, current generative systems disrupt this process as minor prompt changes often yield disproportionate transformations, forcing users to restart from scratch rather than refine an existing design.

These challenges reflect deeper issues identified in interaction design theory as the \textit{gulf of execution} (the distance between user intentions and available actions) and the \textit{gulf of evaluation} (the effort required to interpret system state and determine if goals were achieved)~\cite{hutchins1986direct}.
In generative UI systems, both gulfs are particularly wide: users cannot easily express their design intentions through prompts (execution), nor can they readily understand how their prompts influenced the generated output (evaluation). Moreover, these gulfs amplify each other through iteration.

To address these challenges, we conducted a systematic investigation into how design intent can be effectively communicated to and interpreted by generative AI systems.
Our approach began with a thematic analysis of prompting guidelines from leading UI generation services, through which we identified and categorized the semantic elements essential for UI generation.
This analysis revealed that UI generation semantics operate across multiple levels of abstraction, from high-level product vision to specific component properties.
Building on this framework, we developed a semantic-based UI generation system that bridges the gulfs between user intent and AI output.
Rather than treating prompts as direct instructions, our approach introduces an intermediate semantic layer that serves as a bridge between natural language and generated designs.
These semantics enable bidirectional interaction—both parsing user intentions into structured representations and extracting implemented semantics from generated outputs.
We expose the intermediate semantic layer that connects what users want to what the system generates, allowing users to see and control how their design intentions are composed and translated into actual UI elements.
Furthermore, the system reveals semantic relationships, helping users identify conflicts and maintain coherence.
This transparency enables what was previously a black-box operation to become an inspectable and controllable pipeline.

This work makes three primary contributions:
\begin{itemize}
    \item We provide a hierarchical framework of design semantics for UI generation.
    \item We demonstrate how exposing semantic representations bridges the gulf between user intent and generated output.
    \item We present a comparative study showing that semantic-based generation enhances users' perceived control over intent expression, output interpretation, and iterative refinement.
\end{itemize}

Our semantic-based approach suggests a shift from trial-and-error prompting to systematic, controllable design generation---making the implicit explicit and the opaque transparent.

%% file: section/2-relatedwork.tex
\section{Related Work}

\subsection{UI Generation}

Early work on automatic UI creation was largely rule-based, treating interface synthesis as an optimization over explicit tasks, widgets, and device constraints.
For example, SUPPLE~\cite{gajos2004supple} formalized UI generation as decision-theoretic optimization to personalize layouts to users and devices.
As data-driven methods matured, researchers began learning to map pixels or structured element sets directly to code and layouts: pix2code~\cite{beltramelli2018pix2code} demonstrated end-to-end UI-to-code from screenshots, and UI-Diffuser~\cite{wei2023boosting} and GANSpiration~\cite{mozaffari2022ganspiration} showed the generation of UI images. GAN-based~\cite{goodfellow2020generative} layout generators such as LayoutGAN~\cite{li2019layoutgan} and diffusion-based~\cite{diffusion} models such as LayoutDM~\cite{chai2023layoutdm} learned distributions over element boxes, alignment, and constraints.

Meanwhile, LLM-based text-to-UI generation tools can produce high-fidelity UIs directly from text descriptions, blurring the line between design specification and implementation.
Recent state-of-the-art UI generation services such as Vercel's v0~\cite{v0}, Google Stitch~\cite{stitch}, and Lovable~\cite{lovable} exemplify this trend. 
These systems are trained using not only visual screen data but actual code implementation, enabling the rapid generation of \textit{working} UI prototypes.
However, this power comes with new challenges: the need to effectively communicate design intent to the generative model, and to ensure the AI’s output truly aligns with the user’s goals.
This is driving research into better intent specification and semantic control within UI generation, as discussed next.

\subsection{Design Intent Specification in Generative AI}

As UI generation shifts to natural-language prompts, practitioners face a significant articulation gap in conveying their intent.
Chen \etal~\cite{genuistudy} observed that UX practitioners iteratively engage in trial-and-error with prompts, spending considerable time reformulating descriptions, and still frequently end up with UIs that miss their envisioned goals.

These problems are commonly associated with prompt-based generative AI.
Zamfirescu-Pereira \etal~\cite{zamfirescu2023johnny} found non-AI-experts routinely failed at end-user prompt engineering despite tool support, highlighting ambiguity and brittle phrasing as core barriers.
To reduce such burdens, systems externalize ``good prompting'' into visual programming or mixed-initiative workflows:
ChainForge~\cite{arawjo2024chainforge} lets users compare prompts through a visual data-flow; Promptify~\cite{brade2023promptify}, PromptMagician~\cite{feng2023promptmagician}, and PromptCharm~\cite{wang2024promptcharm} suggest or retrieve style or subject keywords to guide iterative refinement; and Luminate~\cite{suh2024luminate} structures exploration along explicit dimensions so users vary intent deliberately.
Complementary tools help translate fuzzy ideas into concrete inputs: CreativeConnect~\cite{choi2024creativeconnect} extracts keywords from reference images to steer recombination, while GenQuery~\cite{son2024genquery} concretizes abstract text queries into visual exemplars.
Beyond better text, several projects rethink the interaction style: AI-Instruments~\cite{riche2025ai} embodies prompts as interface objects that can be composed and reused; DirectGPT~\cite{masson2024directgpt} applies direct manipulation that the system translates into engineered prompts; and Misty~\cite{Misty} adopts conceptual blending, which lets developers combine properties from examples while expressing intent at different stages of prototyping. Together, these interfaces shift prompt authoring from one-shot wording to manipulating representations of intent.

Within UI work specifically, researchers also target the evaluation and alignment of intent with generated artifacts.
Duan \etal~\cite{duan2024generating} provide actionable automatic feedback on mockups using LLMs. At the workflow level, studies find UX designers currently see generative AI as assistive rather than substitutive and that those tools can both reduce and introduce collaboration conflicts depending on the quality and transparency of outputs~\cite{li2024user}.
Bridging requirements and prototypes, Kolthoff \etal~\cite{kolthoff2024interlinking} semi-automatically link user stories to GUI elements, detecting unimplemented stories and recommending components.
Across these strands, the through-line is clear: effective intent specification benefits from structured, inspectable intermediates rather than raw text alone.
However, existing approaches typically address either input structuring or output evaluation, and most operate on general-purpose dimensions rather than UI-specific semantics.
Our approach contributes a domain-specific semantic framework derived from multifaceted UI prompting practices, combined with \textit{bidirectional semantic guidance} that both parses user intent into a structured form and extracts how semantics are realized in generated outputs—connecting specification and evaluation within a unified intermediate layer.

\subsection{UI Semantic Representations}

Underpinning both the generative models and the intent specification challenge is the notion of UI semantics---ways to represent what a user interface means (\eg, structure, content, and behavior) rather than just how it looks pixel-wise.

Model-based user interface (MBUI) approaches have employed intermediate representations to bridge design intent and implementation~\cite{uipilot}.
Early systems separated dialog content from look-and-feel specifications~\cite{vanderzanden1990jade}, introduced assistance for presentation design~\cite{kim1993highlevel}, and framed UI generation as optimization over abstract specifications~\cite{gajos2004supple}. However, these approaches relied on manually authored models.

More recent data-driven work has developed representations that encode UIs in terms of their constituent elements and relationships, enabling computational analysis and generation.
The Rico dataset~\cite{deka2017rico, rico-semantic} made this concrete at scale with screens containing view hierarchies, text, and interaction traces, seeding work that treats a UI as a composition of elements and constraints rather than a bitmap.
Building atop such corpora, Screen2Vec~\cite{screen2vec} learns embeddings from screens and components from hierarchy context and usage.
Complementing embedding approaches, Graph4GUI~\cite{jiang2024graph4gui} represents each UI as a graph that captures visuo-spatial and semantic relations between elements, improving layout understanding with alignment and containment constraints.
Parallel efforts align vision and language semantics to make quality and intent measurable: UIClip~\cite{wu2024uiclip} scores a UI screenshot against a natural-language description, providing a data-driven signal of design quality/fit that can steer generation and exemplar search.
Recent fine-tuned VLMs, such as ILuvUI~\cite{jiang2025iluvui}, have been instruction-tuned to suggest descriptions, QA, and multi-step planning, broadening the kinds of semantic questions one can ask of an interface. 

Beyond single screens, newer work elevates task-level semantics so that intent and behavior remain traceable across flows. Park \etal~\cite{sparkICML} formulate app-level retrieval and assess design consistency. S\&UI~\cite{s&ui} extracts multi-level UI semantics for UI inspiration search, indicating that explicit semantics help designers find and reason about options at the level they think.
Recent work has also explored intermediate representations for generative UIs, including task-driven data models~\cite{cao2025generative} and malleable overview-detail interfaces~\cite{min2025malleable,min2025meridian}.
While these advances address task structure and pattern-level customization, they do not explicitly model the hierarchical relationships between design intents across abstraction levels.
Our approach addresses this gap from a complementary angle: by analyzing prompting guidelines from existing UI generation services, we derive a semantic framework that structures hierarchical, interdependent information for generation, providing an inspectable intermediate that links designer intent to concrete UI decisions.

%% file: section/3-gulfs.tex
\section{Gulfs in Current UI Generation}
\label{sec:gulf}

While generative AI has significantly advanced capabilities, transforming UI creation from direct manipulation to natural language interaction, this shift has introduced fundamental challenges that mirror and extend Norman's gulfs of execution and evaluation~\cite{hutchins1986direct, gulfOfEnvisioning}.
Through the analysis of a recent comprehensive study on UI generation usage~\cite{genuistudy}, we identify how these interaction challenges manifest themselves in text-to-UI systems and, critically, how they amplify each other through iterative refinement.

\subsection{Gulf of Execution: \\The Articulation Challenge}

The gulf of execution in UI generation manifests as designers' inability to determine what to express and how to articulate it in prompts.
Unlike traditional design tools (\eg, Figma~\cite{figma}), where actions directly manipulate visual elements, generative systems require translating multidimensional design intentions—visual aesthetics, functional requirements, and user contexts—into linear text sequences. 
Users face two fundamental uncertainties: identifying which design dimensions require explicit specification, and finding a suitable way to prompt concepts that often lack precise verbal equivalents.
For instance, when designing an e-commerce checkout page, the user must decide which elements to specify, how visually appealing they should be, and how to induce purchase.
Even when users know what they want, conceptual descriptors such as ``modern,'' ``professional,'' and ``trustworthy'' carry vastly different interpretations: should ``trustworthy'' translate to blue colors (associated with trust), serif typography (traditional authority), or increased whitespace (clarity)?

The GenUI study~\cite{genuistudy} revealed that participants consistently struggled with this articulation challenge, spending significant time searching for how to prompt and attempting multiple reformulations before achieving even partially satisfactory results.
This pattern indicates not a learning curve issue but a fundamental limitation of unstructured text input for conveying complex, hierarchical design intentions.
The absence of systematic input mechanisms forces users into a trial-and-error process where success depends more on chance discovery of effective phrasings than on design expertise, suggesting a critical need for structured approaches to design specification.

\subsection{Gulf of Evaluation: \\The Interpretation Challenge}
The gulf of evaluation emerges when designers cannot understand how their inputs are translated into specific design decisions or assess whether the generated output appropriately reflects their intentions.
When AI systems produce complete interfaces from brief prompts, they make numerous implicit decisions about layout, colors, and interaction patterns, none traceable to specific prompt elements.
For example, if a ``professional dashboard'' prompt yields a dark theme, users cannot determine whether this reflects the AI's interpretation of ``professional,'' a common dashboard pattern, or an arbitrary aesthetic choice.
This opacity transforms evaluation from a deliberate assessment process into detective work, where users have to reverse-engineer the AI's interpretation.
The challenge intensifies when outputs appear visually polished but may not align with unstated functional requirements or design principles, making it difficult to distinguish between acceptable alternatives and fundamental misunderstandings of intent.

Participants in the GenUI study reported spending more time trying to understand what was generated than actually refining designs, indicating a significant evaluation burden that traditional tools don't impose.
The black-box nature of generation creates a fundamental disconnect between designer intent and system output, highlighting the need for mechanisms that explicitly connect high-level requests with their low-level implementations.

\subsection{The Amplification Problem: \\Gulfs Compounding through Iteration}

UI design is inherently iterative, yet in generative systems, iteration amplifies rather than resolves the twin gulfs, creating a compounding problem that degrades design coherence over time.
Each refinement cycle begins with an imperfect understanding of the generated output (evaluation gulf), leading to imprecise modification requests (execution gulf), which produce even less predictable results—a vicious cycle where uncertainty multiplies rather than diminishes.
This amplification manifests as ``semantic drift,'' where successive modifications gradually pull the design away from its original intent, with each iteration introducing new interpretations that conflict with established patterns.
While current AI systems maintain conversation history within a chat as context, they struggle to preserve design principles and relationships across iterations.
Each new prompt can override previous decisions without warning, and competing directives accumulate without resolution—``minimalist but engaging,'' ``professional yet approachable''—creating muddy compromises that satisfy neither goal.

Karnatak \etal~\cite{karnatak2025expandinggenerativeaidesign} found that designers often abandon iterative refinement after a few attempts, concluding that starting over was more efficient than trying to fix degraded designs.
This breakdown of the iterative design process represents a fundamental failure of current generative systems to support the exploratory, refinement-based workflow that characterizes professional design practice.
These compounding gulfs reveal a fundamental need for an intermediate representation—one that can structure design intentions, make AI decisions interpretable, and maintain consistency across iterations.

%% file: section/4-semantics.tex
\section{Understanding UI Generation Semantics}
\label{sec:semantic}

To address the gulfs identified in~\Cref{sec:gulf}, we first need to understand what semantics are required and how they are structured for effective UI generation.
We analyzed prompting guidelines and best practices from major UI generation services to identify common, effective patterns in how users specify UI design intent.

\begin{figure*}
    \centering
    \includegraphics[width=\linewidth]{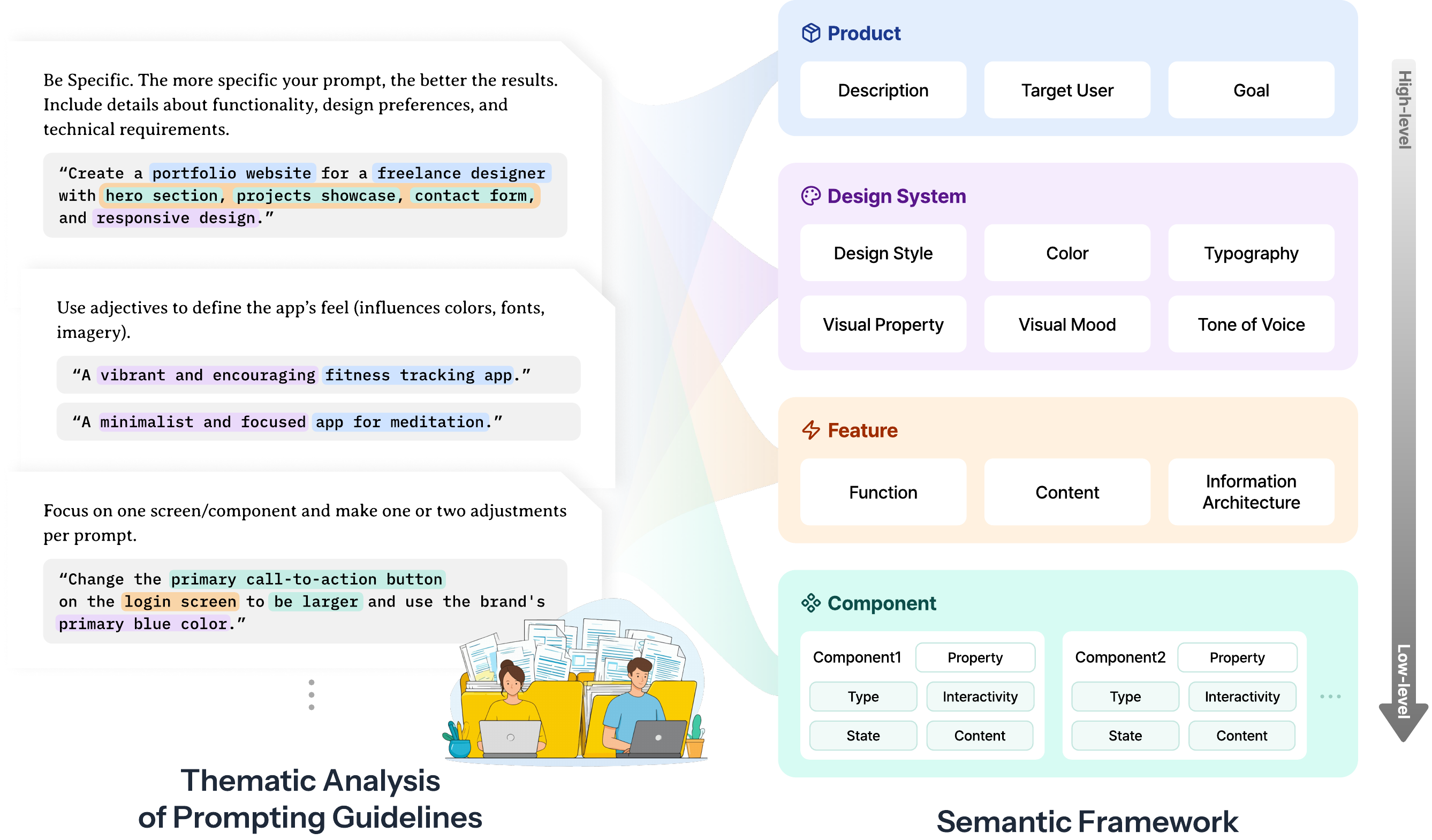}
    \caption[]{\textit{From prompting guidelines to a semantic framework.} \textmd{We conducted a thematic analysis of prompting guidelines from major UI generation tools, surfacing recurring patterns of what information users specify. This yields a four-level hierarchical representation — \ProductTag{\raisebox{-0.15em}{\includegraphics[height=0.9em]{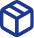}}\:Product}, \DesignSystemTag{\raisebox{-0.15em}{\includegraphics[height=0.9em]{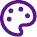}}\:Design System}, \FeatureTag{\raisebox{-0.15em}{\includegraphics[height=0.9em]{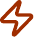}}\:Feature}, \ComponentTag{\raisebox{-0.15em}{\includegraphics[height=0.9em]{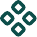}}\:Component} — that organizes interdependent design semantics from high to low levels. Semantic elements are related both vertically (between levels) and horizontally (within levels), meaning that changes to one element can cascade across others.}}
    \Description{Figure showing how prompting guidelines from UI generation tools were analyzed into a four-level semantic framework. Thematic examples of prompts on the left are mapped to hierarchical categories on the right: Product, Design System, Feature, and Component.}
    \label{fig:thematic}
\end{figure*}

\subsection{Analysis Process}
We collected prompting guidelines with example prompts, which contain best practice recommendations, from six leading prompt-based UI generation services (Vercel v0~\cite{v0}, Google Stitch~\cite{stitch}, Figma Make~\cite{figmamake}, Uizard~\cite{uizard}, Lovable~\cite{lovable}, and Relume~\cite{relume}), resulting in 907 guide fragments consisting of 601 guidelines and 306 prompt examples.
Three authors collaboratively analyzed these materials through iterative coding, identifying recurring patterns and categorizing the types of information required for different UI generation tasks.
We conducted exploratory analysis, allowing patterns to emerge from the data while maintaining a focus on what information users need to provide and how it relates to generation outcomes.

\subsection{Hierarchical Semantic Framework}

Our analysis revealed that UI generation encompasses multiple task types and scopes.
We identified three primary action types: creation (generating new UI from scratch), modification (altering existing designs), and analysis (extracting or evaluating design properties).
These actions operate at different scope levels: entire applications, partial sections or screens, and individual elements.
Each combination of action type and scope requires different semantic information, and effective UI generation must coordinate this information across multiple abstraction levels, where changes at one level can influence other levels.
Across these varying tasks and scopes, the semantic information organizes into four hierarchical levels: \ProductTag{\raisebox{-0.15em}{\includegraphics[height=0.9em]{figures/icons/product_icon.pdf}}\:Product},~\DesignSystemTag{\raisebox{-0.15em}{\includegraphics[height=0.9em]{figures/icons/system_icon.pdf}}\:Design System},~\FeatureTag{\raisebox{-0.15em}{\includegraphics[height=0.9em]{figures/icons/feature_icon.pdf}}\:Feature},~and~\ComponentTag{\raisebox{-0.15em}{\includegraphics[height=0.9em]{figures/icons/comp_icon.pdf}}\:Component}.
While this structure resembles Garrett's layers of user experience \cite{garrett2010elements} and the visual-building logic of Atomic Design~\cite{frost2016atomic}, our framework encompasses both perspectives, spanning from high-level experience to specific elements.
In doing so, it captures the broad semantics expressed in generative workflows and the relationships that shape how those semantics influence generated outcomes.

\leavevmode\ProductTag{\raisebox{-0.15em}{\includegraphics[height=0.9em]{figures/icons/product_icon.pdf}}\:Product} encompasses three key elements: Description (what is being built), Target User (who will use it), and Goal (why it exists).
This level establishes the overarching context that influences all other design decisions.
These high-level decisions provide critical context that shapes interpretation at lower levels.

\leavevmode\DesignSystemTag{\raisebox{-0.15em}{\includegraphics[height=0.9em]{figures/icons/system_icon.pdf}}\:Design System} defines the visual and experiential language through six main categories: Design Style (overall aesthetics; \eg, minimalist, skeuomorphic, or glassmorphic), Color (palette and scheme), Typography (fonts, their sizes and hierarchies), Visual Properties (\eg, shadows, corner radius, spacing), Visual Mood (emotional tone; \eg, warm), and Tone of Voice (UX writing style).
These elements serve as cross-cutting concerns~\cite{kiczales1997aspect}, ensuring consistency across all features and components.

\leavevmode\FeatureTag{\raisebox{-0.15em}{\includegraphics[height=0.9em]{figures/icons/feature_icon.pdf}}\:Feature} addresses screen-specific requirements through three components: Function (what the screen accomplishes), Content (the information to display), and Information Architecture (how the content is structured and prioritized).
This level bridges high-level product goals with concrete component implementations.

\leavevmode\ComponentTag{\raisebox{-0.15em}{\includegraphics[height=0.9em]{figures/icons/comp_icon.pdf}}\:Component} represents the most concrete specifications: Type (\eg, semantic HTML elements~\cite{Semantic89:online} such as buttons, forms, navigation), Interactivity (\eg, hover states, click behaviors, transitions, or animations), State (\eg, active, disabled, loading conditions), Content (actual text or data), and Properties (component-specific attributes).
Notably, elements like interactivity and state—often absent in static UI taxonomies—emerged as meaningful semantics in generative contexts, indicating that generative workflows surface design dimensions that are otherwise latent.

\subsubsection*{Example Walkthrough}
To illustrate the hierarchical relationships of these semantics, consider a music streaming product such as Spotify.
At the \ProductTag{\raisebox{-0.15em}{\includegraphics[height=0.9em]{figures/icons/product_icon.pdf}}\:Product} level, defining the Goal as ``providing a personalized and immersive audio experience'' for a Target User who values ``effortless connection to music'' sets the foundational context.
This intent relates to the \DesignSystemTag{\raisebox{-0.15em}{\includegraphics[height=0.9em]{figures/icons/system_icon.pdf}}\:Design System}: to support immersion, the Color is defined as ``Dark Mode'' to make album artwork serve as the primary visual focus, while the Visual Mood is set to ``energetic and bold'' to guide stylistic choices.
Descending to the \FeatureTag{\raisebox{-0.15em}{\includegraphics[height=0.9em]{figures/icons/feature_icon.pdf}}\:Feature} level, exemplified by the home screen of the music app, the Function of ``personalized recommendation hub'' drives the Information Architecture to organize content into ``horizontal scrollable feeds'' rather than vertical lists, prioritizing visual scanning over textual density.
Finally, at the \ComponentTag{\raisebox{-0.15em}{\includegraphics[height=0.9em]{figures/icons/comp_icon.pdf}}\:Component} level, these constraints materialize in specific Interactivity and Properties: an Album Card is implemented not merely as an image, but with a ``hover-triggered Play button'' and ``rounded corners,'' serving the high-level product goal of reducing friction for playback.

In addition to these top-down influences, semantics also interact \textit{horizontally} within levels to ensure coherence.
For instance, within \DesignSystemTag{\raisebox{-0.15em}{\includegraphics[height=0.9em]{figures/icons/system_icon.pdf}}\:Design System} level, the ``energetic'' Visual Mood necessitates a vibrant Color palette (Neon Green) rather than muted pastels; a mismatch here would create a semantic conflict.
Similarly, at the \FeatureTag{\raisebox{-0.15em}{\includegraphics[height=0.9em]{figures/icons/feature_icon.pdf}}\:Feature} level, the high-fidelity Content (album artwork) horizontally influences the Information Architecture to adopt a grid-based layout instead of a dense text list, ensuring that the content type and structural arrangement are logically aligned.

\subsection{Key Insights}
As illustrated in the walkthrough, critical characteristics of our framework analysis are the relationships between semantic elements.
These relationships operate both \textit{vertically} (between levels) and \textit{horizontally} (within levels).
Understanding these relationships is essential because changes to one semantic element can have far-reaching effects throughout the design in the generation context.
For example, changing the target user from ``young professionals'' to ``older adult users'' doesn't just affect font size; it influences color contrast, interaction patterns, information density, and navigation complexity.
Together, these patterns indicate that prompt-based UI generation requires coordinating meaning across multiple abstraction layers rather than simply describing visual appearance—helping explain why current text-based prompting feels difficult and unpredictable.
Users must somehow encode this multi-level, interconnected information into linear text, without clear guidance on what to specify explicitly versus what to leave for UI generation.

Our analysis demonstrates that effective UI generation requires more than just describing visual appearance or listing desired features.
It demands understanding and specifying information across multiple abstraction levels while managing the complex relationships between them.
The current approach of unstructured text prompts cannot adequately capture this semantic richness, nor can it make the relationships between different design decisions explicit.
This framework analysis provides the foundation for designing systems that can better support users in specifying, understanding, and refining generated UI designs.

%% file: section/5-system.tex
\section{System}

To bridge the gulfs identified in~\Cref{sec:gulf}, we designed and implemented a system that leverages the semantic framework from~\Cref{sec:semantic} as an intermediate representation between human intent and AI-generated user interfaces.
We developed a web-based application that provides structured mechanisms for specifying, analyzing, and refining UI designs through explicit semantic manipulation.

\subsection{Design Goals}
We define three design goals aligned with the gulfs identified in~\Cref{sec:gulf}:
\begin{enumerate}
    \item[\textbf{DG1}] \textbf{Structured Specification} addresses the gulf of execution by replacing ambiguous prompts with a hierarchical semantic form. Rather than struggling to articulate design intent in natural language, users specify their requirements through a hierarchical framework that guides what information is needed at each level. This structured approach reduces the guesswork of prompt engineering while maintaining flexibility through partial specification.
    \item[\textbf{DG2}] \textbf{Transparent Analysis} tackles the gulf of evaluation by extracting and presenting the semantic decisions that the model enacted in the generated UI. Instead of treating AI generation as a black box, the system analyzes the output to reveal what design choices were made and how they relate to user input. This transparency enables users to understand how their intents were actually implemented.
     \item[\textbf{DG3}] \textbf{Relationship-aware Refinement} prevents the amplification problem by maintaining semantic relationships throughout iterations. By visualizing dependencies and potential cascades, the system helps users understand how modifications will propagate through the design and mitigates the coherence degradation that typically occurs in iterative refinement.
\end{enumerate}

\begin{figure*}
    \centering
    \includegraphics[width=0.94\linewidth]{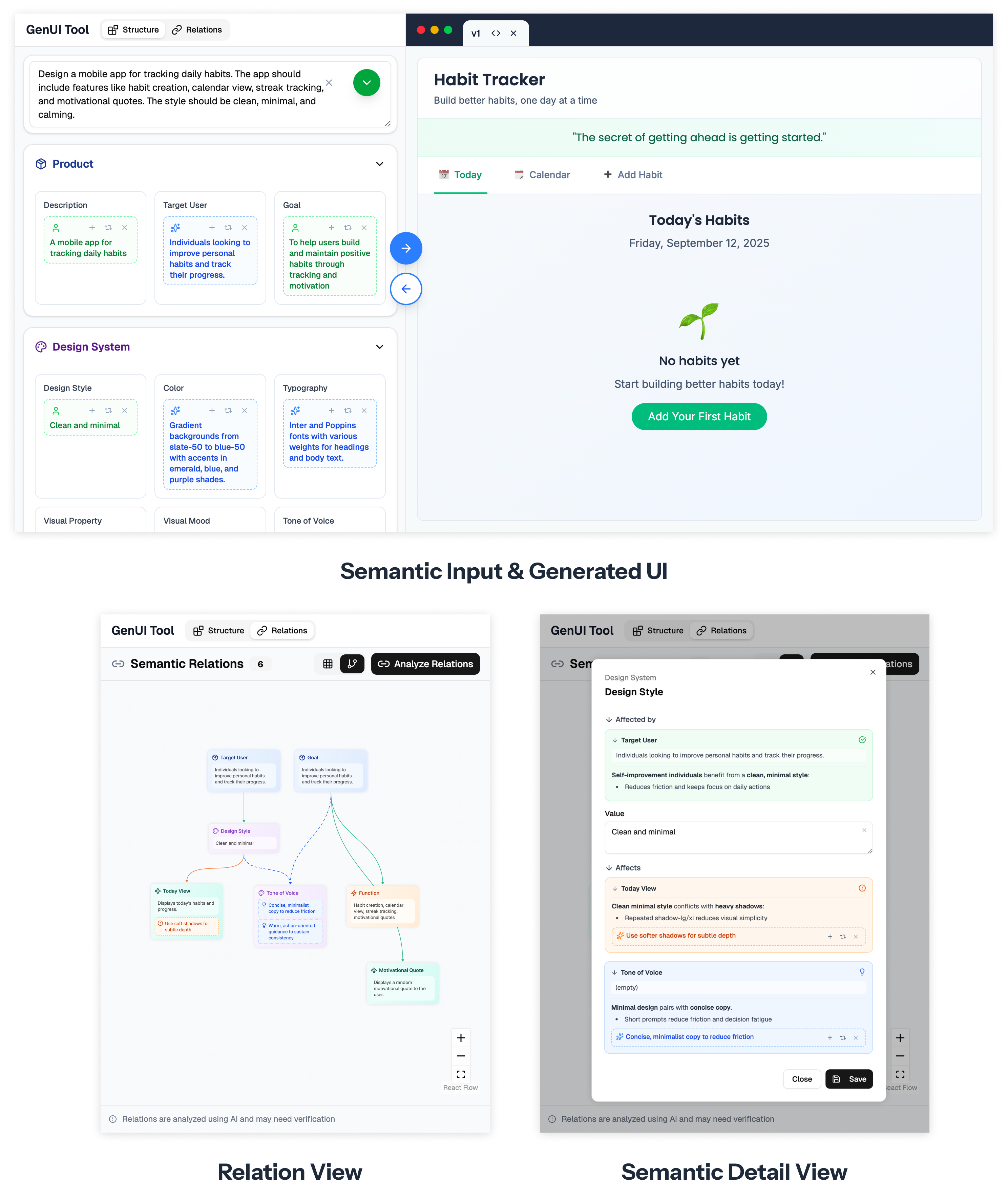}
    \caption{\textit{System interface with semantic input, generated UI, and analysis views.} \textmd{Top: Users specify semantics across the four-level framework (left), which are used to generate the corresponding UI (right). Bottom: The Relation View visualizes dependencies among semantic attributes, while the Semantic Detail View shows a selected attribute with its values, upstream and downstream relations, and suggested refinements for coherent design decisions.}}
    \Description{System interface showing how user-specified semantics generate a corresponding UI and support analysis. The top panels display semantic input on the left and the generated habit tracker UI on the right. The bottom panels show a Relation View of dependencies among semantics and a Semantic Detail View explaining one attribute with its related values.}
    \label{fig:system-overview}
\end{figure*}

\begin{figure*}
    \centering
    \includegraphics[width=\linewidth]{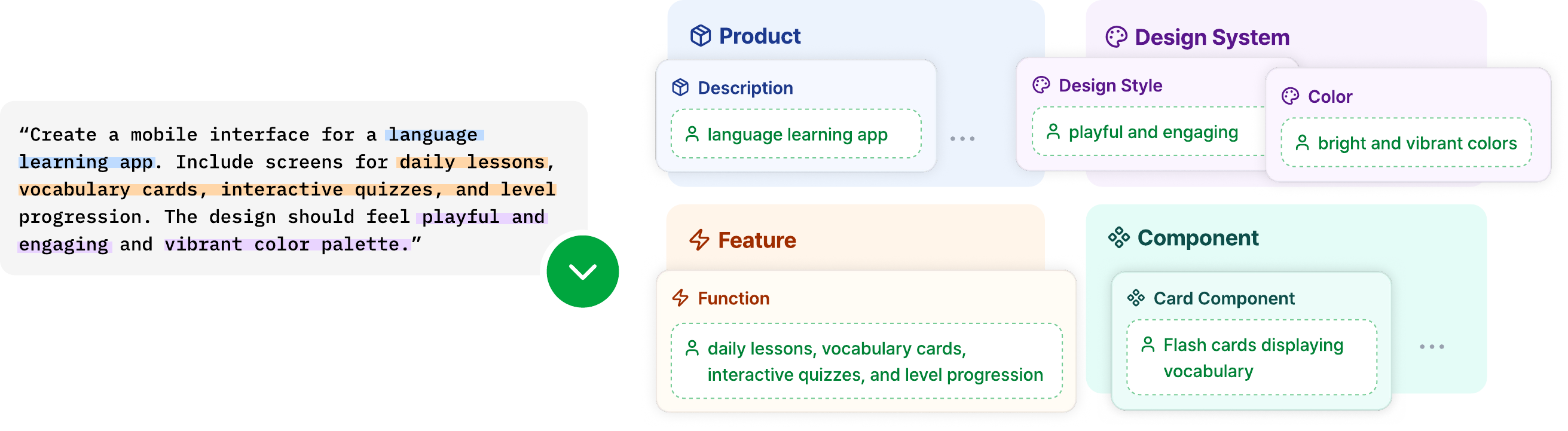}  
    \caption{\textit{Example of input semantics specification.} \textmd{
Users can define semantics manually or via natural language prompts. Natural language input is \raisebox{-0.15em}{\includegraphics[height=0.9em]{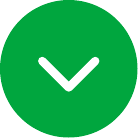}} parsed into structured semantics across our four-level framework: \ProductTag{\raisebox{-0.15em}{\includegraphics[height=0.9em]{figures/icons/product_icon.pdf}}\:Product}---Description, \DesignSystemTag{\raisebox{-0.15em}{\includegraphics[height=0.9em]{figures/icons/system_icon.pdf}}\:Design System}---Design Style and Color, \FeatureTag{\raisebox{-0.15em}{\includegraphics[height=0.9em]{figures/icons/feature_icon.pdf}}\:Feature}---Function, and \ComponentTag{\raisebox{-0.15em}{\includegraphics[height=0.9em]{figures/icons/comp_icon.pdf}}\:Component}---Card Component.}}
\Description{Example of input semantics specification. A natural language prompt for a mobile learning app is parsed into structured categories—Product, Design System, Feature, and Component—making design intent explicit and organized.}
    \label{fig:parsing}
\end{figure*}

\subsection{System Overview}
The system implements a three-phase workflow---Generate, Analyze, and Refine---with semantic structure serving as the persistent intermediate representation throughout.

As shown in~\Cref{fig:system-overview}, users first specify semantics in the semantic structure view across the four-level framework (top-left); the system then generates the corresponding UI from these semantics and renders it (top-right).
This pairing of semantic input and visual feedback anchors user intent to concrete outcomes, reducing ambiguity.
Users can switch to the relation view (bottom-left), which visualizes interdependencies among semantics.
From either view, clicking any semantic attribute opens the semantic detail panel for in-depth inspection of that attribute and its relationships.

Each phase operates independently yet contributes to a cohesive workflow: \raisebox{-0.15em}{\includegraphics[height=0.9em]{figures/icons/gen_right.pdf}} the Generate phase scaffolds design intent, \raisebox{-0.15em}{\includegraphics[height=0.9em]{figures/icons/analyze_left.pdf}} the Analyze phase extracts augmented semantics to make AI reasoning interpretable, and the Refine phase uses relationship analysis to guide iterative improvement.
Together, these views form a feedback loop in which users articulate, observe, and iteratively refine design intent while preserving semantic coherence.

\subsection{Input Semantics Specification}

The input interface implements the four-level semantic framework as collapsible panels, each containing relevant attribute fields.
Users can directly fill in semantic fields for precise control or provide natural language input through the free-form text field, as shown in~\Cref{fig:parsing}, which \texttt{GPT-5}~\cite{gpt-5} parses into structured semantics.
Critically, we designed the system to allow partial specification—users need only provide information they consider important, letting the AI infer the remaining details.
The structured semantics are compiled into markdown format and sent to the UI generation model (Vercel \texttt{v0-1.5-md}) to produce React component code, with the real-time preview shown alongside the semantic specification.

\begin{figure*}[!htb]
    \centering
    \includegraphics[width=\linewidth]{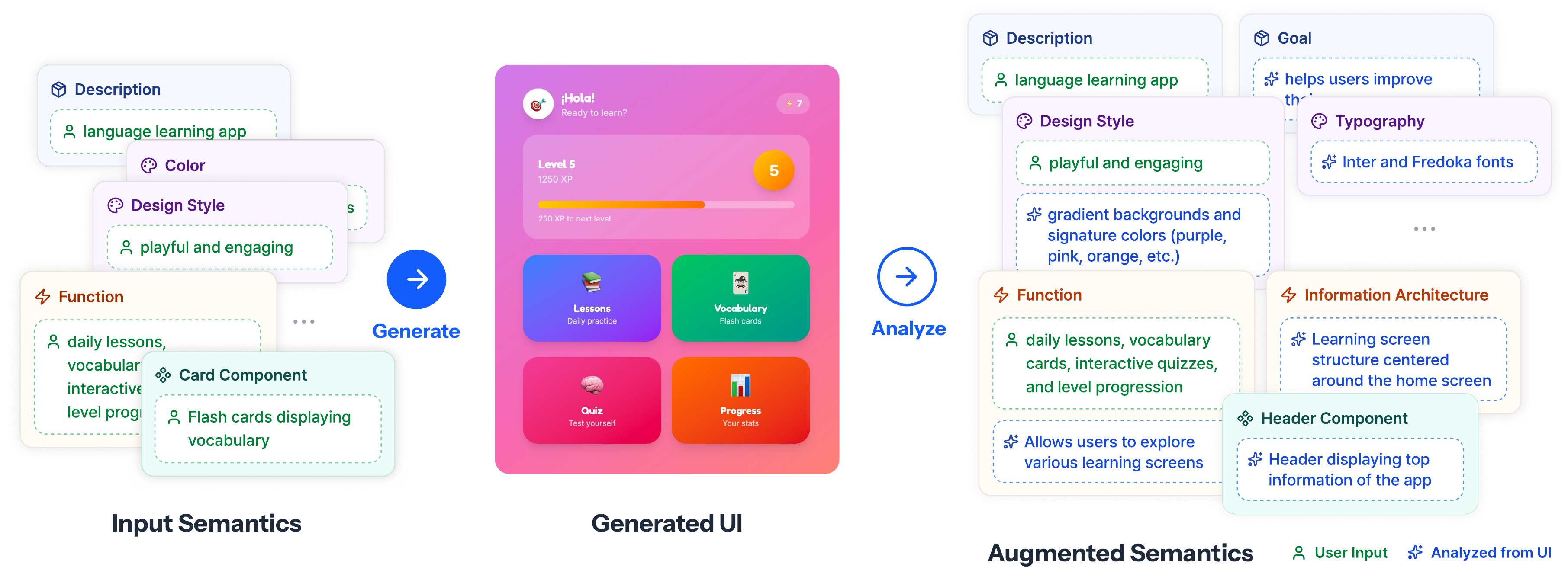}
    \caption{\textit{System workflow from input semantics to augmented semantics.} \textmd{Input semantics provided by the user (left) are transformed into a functional UI through the \raisebox{-0.15em}{\includegraphics[height=0.9em]{figures/icons/gen_right.pdf}} Generate step (center).
    When \raisebox{-0.15em}{\includegraphics[height=0.9em]{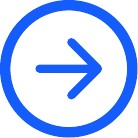}} Analyze is triggered, the system augments this view with additional semantics inferred from the UI (right).}}
    \Description{Workflow diagram showing how user-specified input semantics are turned into a generated UI and then analyzed to produce augmented semantics. The figure illustrates that initial inputs like description and design style generate a UI, while the analysis step adds inferred details such as typography and information architecture.}
    \label{fig:generate-analyze}
\end{figure*}

\subsection{Augmented Semantics Analysis}

Upon generation, users can trigger the \raisebox{-0.15em}{\includegraphics[height=0.9em]{figures/icons/analyze_left.pdf}} Analyze function to extract semantic information from the generated UI code. \texttt{GPT-5}~\cite{gpt-5} examines the code and the corresponding screenshot of the generated UI to identify implemented design patterns, visual properties, and structural decisions.
This extraction process reveals not only what was explicitly requested but also what the AI inferred or added based on its interpretation of the input semantics.

\Cref{fig:generate-analyze} shows how the system enhances transparency through augmented semantics.
When the Analyze is triggered, the system extracts not only what was explicitly specified (\eg, ``playful and engaging'' style) but also what was inferred or added (\eg, typography, layout hierarchy, or interaction flow).

The analysis results appear as \textit{Augmented Semantics}—a comprehensive semantic representation of the actual implementation.
Presented in the semantics panel with blue highlights, this comparison makes the AI's interpretation process transparent. 
For instance, if a user specified ``playful and engaging'' as the Design Style, the augmented semantics might reveal this was implemented in Color through ``gradient backgrounds and signature colors (purple, blue, pink, orange)'' along with ``rounded corners'' and ``bold typography.''
This explicit mapping between abstract intent and concrete implementation addresses the interpretation challenge users face with traditional generation systems.

\begin{figure*}
    \centering
    \includegraphics[width=\linewidth]{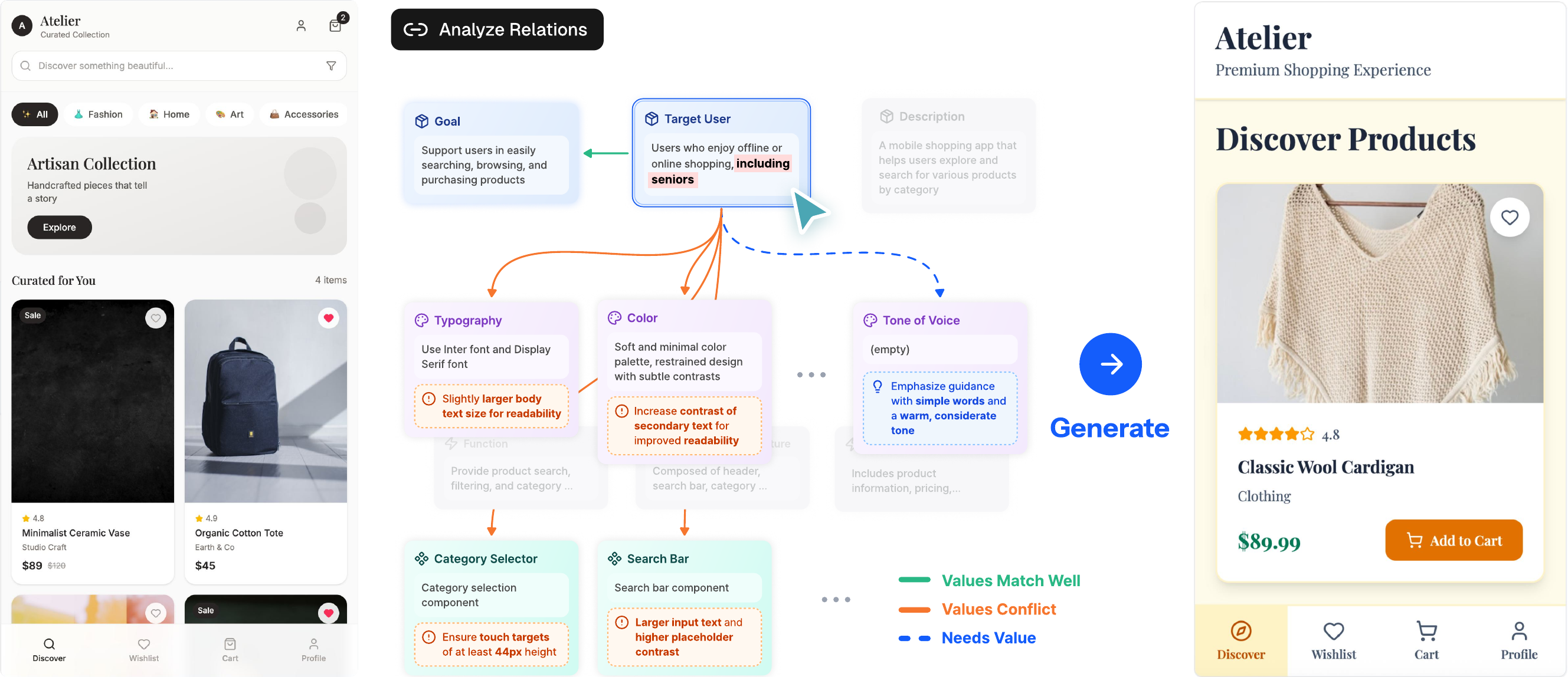}
    \caption{\textit{Example of semantic relationship analysis.} \textmd{A mobile shopping app specification is analyzed to reveal dependencies between semantic attributes. The system visualizes relations as a graph, where arrows denote influences between attributes: green edges \raisebox{-0.15em}{\includegraphics[height=0.9em]{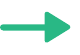}} indicate Values Match Well, orange edges \raisebox{-0.15em}{\includegraphics[height=0.9em]{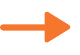}} highlight Values Conflict, and blue dashed edges \raisebox{-0.15em}{\includegraphics[height=0.9em]{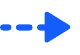}} denote Needs Value.}}
    \Description{Example of semantic relationship analysis for a shopping app. The system shows a graph of semantic attributes such as Target User, Typography, Color, and Search Bar, with arrows indicating influences. Green edges mark matches, orange conflicts, and blue needs. The generated UI on the right reflects these recommendations, like larger text and higher contrast.}
    \label{fig:relation}
\end{figure*}

\subsection{Semantic Relationship Analysis}

As shown in~\Cref{fig:relation}, the relationship analysis renders a relation graph between semantic attributes.
The graph represents semantics as nodes and their relationships as edges, with three relationship types: \textit{Values Match Well} \raisebox{-0.15em}{\includegraphics[height=0.9em]{figures/icons/relation_green.pdf}} in green, \textit{Values Conflict} \raisebox{-0.15em}{\includegraphics[height=0.9em]{figures/icons/relation_red.pdf}} in orange, and \textit{Needs Value} \raisebox{-0.15em}{\includegraphics[height=0.9em]{figures/icons/relation_blue.pdf}} in blue dashed lines.
This graph serves dual purposes: guiding input decisions before generation and explaining implementation patterns after generation.

In a pre-generation context, relationship analysis helps users understand the implications of their specifications.
When a user sets ``elderly users'' as the Target User, the system can suggest related semantics like ``larger typography,'' ``high contrast colors,'' and ``simplified navigation.''
These suggestions provide guidance on coherent design decisions.
The system also identifies potential conflicts—for example, specifying both ``information-dense'' and ``minimalist'' triggers a warning about contradictory requirements.

Post-generation relationship analysis reveals how AI interpreted and connected different semantic elements.
Clicking on any semantic node opens a detailed view showing \textit{Affected by} relationships (which semantics influenced this one) and \textit{Affects} relationships (which semantics this one influences).
Each relationship includes an explanation, such as ``Color palette was influenced by Target User (elderly) to ensure sufficient contrast for readability.''
This transparency helps users understand the design logic and make informed refinement decisions.

\subsection{Iterative Refinement Support}
The system addresses the amplification problem through a dual-persistence strategy that anchors both semantic intent and code implementation.
Unlike chat-based systems, where context is treated as passive conversation history prone to drift, our approach leverages explicit semantic diffs to drive targeted regeneration, utilizing the existing code as an active structural constraint.

First, the system preserves the semantic state across iterations and utilizes it in every generation.
Second, to support granular, section-specific modifications, we designed a targeted regeneration strategy.
While standard chat interfaces rely on the model to implicitly infer changes from a stream of conversation history, our system explicitly specifies the semantic diff (comparing previous semantics and identifying which semantic value changed and how) in the prompt.
Given that the model regenerates the entire component, by providing previously generated UI code as a reference, the system instructs the model to apply changes to the corresponding modified semantics while preserving the existing structure.
This strategy enables a controlled, section-specific update, preventing unintended side effects even when regenerating the full component.

\subsection{Implementation Details}

We implemented the system as a web-based application using the Next.js framework. The architecture is designed as a modular agentic workflow where the parsing, generation, and analysis functions operate independently to iteratively refine a React component.
Central to this design is the management of semantic data as structured objects. This object-based state serves as the single source of intent, synchronizing the distinct modules and ensuring that design intent is preserved without loss across the Generate-Analyze-Refine loop.

For semantic reasoning tasks, including parsing, relationship analysis, and generated UI analysis, we utilized \texttt{GPT-5}. To guarantee output format consistency, we enforced structured outputs with JSON schemas, preventing formatting errors and hallucinated attributes often found in unstructured LLM responses. The analysis module uniquely leverages \texttt{GPT-5}'s multimodal capabilities, ingesting both the generated React code and rendered screenshots to infer implicit visual properties (\eg, mood, spacing) that are visible in the browser but abstract in the code.

For the generation pipeline, the system interfaces with Vercel's \texttt{v0-1.5-md} API. The system dynamically saves the response code as React components for rendering. To ensure a cohesive and expressive design space, the runtime environment enforces specific technical constraints, including the use of Tailwind CSS for styling and Google Fonts for typography. Detailed implementation specifications, including system prompts and schemas, are provided in the supplementary material.

%% file: section/6-evaluation.tex
\section{Comparative User Study}

We conducted a controlled user study to evaluate whether our semantic-based system addresses the gulfs identified in~\Cref{sec:gulf}.
The study compared our system against a baseline chat-based interface with prompt enhancement features, examining how structured semantic representation impacts UI generation workflows.
Our evaluation addresses three research questions directly tied to the identified gulfs:
\begin{enumerate}
    \item[\textbf{RQ1}] How does semantic-based prompting with relationship analysis (value recommendation) help users articulate design intent, thereby reducing \textit{the gulf of execution}?
    \item[\textbf{RQ2}] How does semantic extraction and relationship analysis (showing matches and conflicts) help users understand generated outputs, thereby reducing \textit{the gulf of evaluation}?
    \item[\textbf{RQ3}] To what extent does the semantic structure make iterative refinement more controllable and predictable, addressing \textit{the amplification problem}?
\end{enumerate}

\begin{table*}[!ht]
  \centering
    \caption{Participant demographics and UI generation experience.}
  \label{tab:participants}
  \begin{tabular}{ccccc|ccccc}
    \toprule
    \textbf{PID} & \textbf{Age} & \textbf{Role} & \textbf{YOE} & \textbf{GenUI Experience} &
    \textbf{PID} & \textbf{Age} & \textbf{Role} & \textbf{YOE} & \textbf{GenUI Experience} \\
    \midrule
   P1  & 26 & SWE & 6  & Expert       & P8  & 20 & UXD & 1 & Beginner \\
P2  & 24 & UXD & 3  & Intermediate & P9  & 22 & PM  & 3 & Beginner \\
P3  & 30 & UXD & 6  & Beginner     & P10 & 33 & UXD & 7 & Expert \\
P4  & 25 & PM  & 3  & Beginner     & P11 & 22 & SWE & 2 & Intermediate \\
P5  & 29 & UXD & 10 & Beginner     & P12 & 29 & SWE & 4 & Expert \\
P6  & 21 & UXD & 3  & Intermediate & P13 & 27 & PM  & 3 & Beginner \\
P7  & 30 & UXD & 4  & Intermediate & P14 & 22 & UXD & 1 & Intermediate \\
\bottomrule
\addlinespace[0.25em]
\multicolumn{10}{c}{\footnotesize YOE = Years of Experience; GenUI = Generative UI; SWE = Software Engineer; UXD = UI/UX Designer; PM = Product Manager.}\\
  \end{tabular}
  \Description{Table summarizing 14 study participants with their age, role, years of experience, and prior generative UI experience. Roles include software engineers, UI/UX designers, and product managers. Experience with generative UI ranged from beginner to expert.}
\end{table*}

\subsection{Participants and Study Design}
We recruited UI/UX practitioners through online design communities, resulting in 14 participants (8 UI/UX designers, 3 software engineers, 3 product managers) with varying levels of experience in UI generation tools (6 beginners, 5 intermediate users, 3 experts).
Participants had an average of 4 years of professional experience (SD = 2.48, range: 1-10 years).
Participants received compensation equivalent to US\$20 upon completing the session.

We employed a within-subjects design where each participant used both systems: our semantic-based interface and a baseline chat-based system with v0's prompt enhancement.
Both systems used an identical UI generation model (\texttt{v0-1.5-md}) to ensure that differences stemmed from the interface design rather than generation capabilities.

Participants completed two types of tasks designed to evaluate different aspects of the UI generation process:
(1) \textit{Target Screen Generation} required participants to create a specific UI based on detailed requirements, testing their ability to express precise design intent and evaluate whether outputs matched specifications.
(2) \textit{Open-ended Screen Generation} allowed creative freedom in designing a UI for a given context, examining how well the systems supported exploratory design and iterative refinement.
They selected UI design scenarios, adopted from Park \etal~\cite{s&ui}, two for each task type, ensuring comparable complexity across conditions. The UI design scenarios covered diverse domains, including e-commerce, education, travel booking, music streaming, and food delivery products.

To mitigate potential learning and carry-over effects, we utilized a counterbalanced within-subjects design. 
Participants were randomly assigned to two groups to alternate the order of system usage.
Regarding task scenarios, participants selected a total of 4 out of 5 available domains to ensure domain familiarity. 
The assignment of these chosen scenarios to each system condition was randomized to prevent repetition and ensure that no scenario was tied exclusively to a specific system.

The 90-minute study session proceeded as follows:
\begin{itemize}
    \item Introduction and Tutorial (10 mins): Participants received training on both systems' features.
    \item Task 1 (30 mins): Target screen generation with both systems, each with a different scenario.
    \item Task 2 (30 mins): Open-ended screen generation with both systems, each with a different scenario.
    \item Post Survey and Interview (20 mins): Participants completed questionnaires evaluating their experience with both systems, followed by a semi-structured interview.
\end{itemize}

For each task, participants first generated an initial UI, then were encouraged to refine the generated UI based on their own design goals.
To ensure sufficient engagement with the iterative capabilities of both systems, participants were instructed to perform a minimum of three iteration cycles per task.
Beyond this requirement, they were free to continue iterating until they reached a state of subjective satisfaction.
This structure allowed us to observe both the gulfs and the amplification problems.

\subsection{Measures}
We collected both quantitative and qualitative data to comprehensively evaluate the systems across our research questions. We adapted a questionnaire from AIChains~\cite{aichains} to measure the overall effectiveness (Match Goal, Think Through, Transparent, Controllable, and Collaborative) of human-AI collaboration using 7-point Likert scales~\cite{likert1932technique}.
This validated instrument captures users' perceived control, understanding, and satisfaction when working with AI systems.
Additionally, we developed a custom questionnaire that aligns with our research questions and our system, measuring Intent Expressiveness, Intent Reflection, Output Interpretability, Ease of Modification, and Modification Reflection.
Likewise, each dimension was measured using multiple items on a 7-point scale (1 = low, 7 = high).
Semi-structured interviews explored participants' experiences in depth, focusing on:
\begin{itemize}
    \item Overall experience and perceived differences between chat-based and semantic-based systems
    \item Intent expression and utility of semantic structure (RQ1)
    \item Result interpretation and value of semantic extraction (RQ2)
    \item Iteration experience and predictability of modifications (RQ3)
    \item System limitations and suggestions for improvement
\end{itemize}

\begin{figure*}[!ht]
    \centering
    \includegraphics[width=\linewidth]{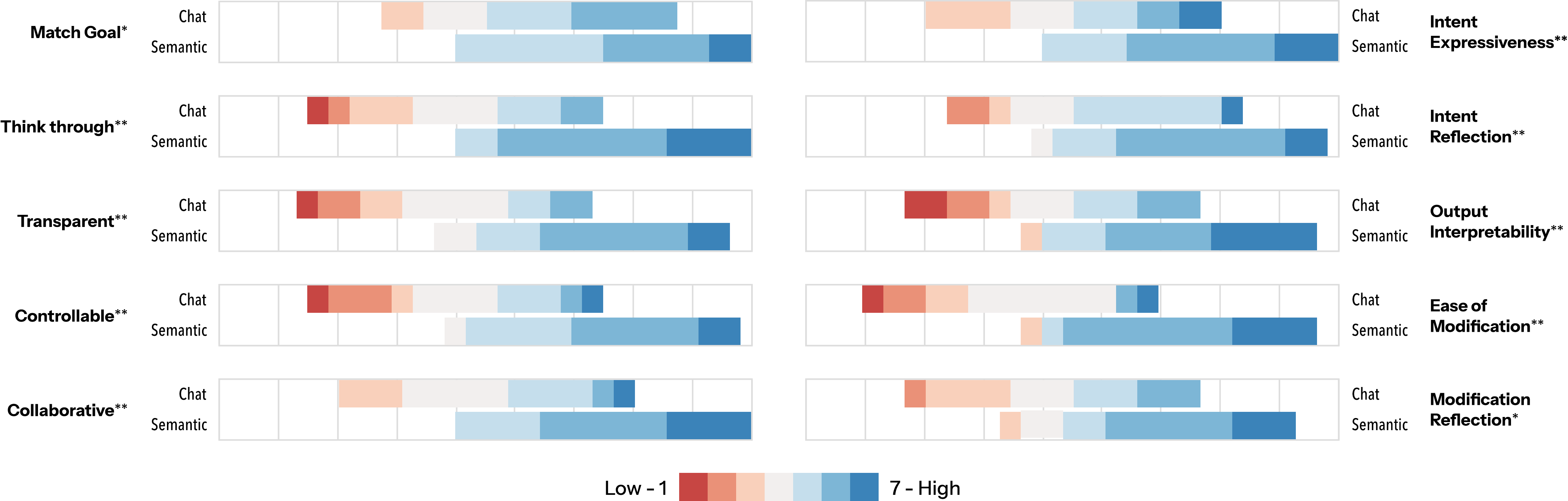}
    \caption{\textit{Comparative user study results across evaluation dimensions.} \textmd{Distribution of 7-point Likert-scale ratings for the baseline chat interface versus our semantic system. Each bar shows participant responses from low (1, red) to high (7, blue). Across all measures, the semantic system received significantly higher ratings than the chat baseline (*$p < .05$, **$p < .01$).}}
    \Description{Comparison of participant ratings between the baseline chat interface and the semantic system across evaluation dimensions. Heatmaps show that the semantic system consistently received higher scores, with responses clustering toward the positive end of the 7-point scale.}
    \label{fig:likert}
\end{figure*}

\subsection{Results}
We present quantitative results from questionnaires followed by qualitative findings from interviews, analyzing the differences in user experience and perceived control between our semantic-based system and the baseline chat interface.

\subsubsection{Quantitative Results}
Given the paired design and Likert-scale nature of our data, we employed Wilcoxon signed-rank tests~\cite{wilcoxon1945individual} to analyze differences between systems.
The tests revealed statistically significant differences across all measured dimensions (*$p<.05$, **$p<.01$) with large effect sizes ($r = Z/\sqrt{N} > .5$).
\Cref{fig:likert} illustrates these differences through distinct response patterns.
The baseline chat interface produced polarized outcomes—users either succeeded or struggled significantly, with little middle ground.
In contrast, our semantic system supported more consistent outcomes, with responses clustering toward positive ratings.

\paragraph{Reducing the Gulf of Execution.}
Participants found our system significantly more effective for expressing design intent.
They rated Intent Expressiveness substantially higher ($p$ = .008, $r$ = 0.79), with semantic system ratings (\texttt{M} = 5.93) notably exceeding the baseline (\texttt{M} = 4.64).
The diverging response patterns in~\Cref{fig:likert} show that while baseline users frequently selected neutral or negative ratings, semantic system users concentrated on positive ratings (6-7 on the Likert scale).
Intent Reflection—how accurately users felt the system understood their intentions—also showed significant improvement ($p$ = .010, $r$ = 0.71; Semantic: \texttt{M} = 5.79, Baseline: \texttt{M} = 4.36).
The Think through metric demonstrated the largest effect ($p$ = .002, $r$ = 0.87), jumping from \texttt{M} = 3.93 in the baseline to \texttt{M} = 6.14 in our system, suggesting that participants perceived the structured semantic input as helpful for systematically considering design decisions rather than struggling with prompt formulation.

\paragraph{Reducing the Gulf of Evaluation.} 
Participants reported that semantic extraction and relationship analysis enhanced their output understanding.
Output Interpretability showed one of the strongest improvements ($p$ = .004, $r$ = 0.82), with mean scores rising from 3.86 (baseline) to 5.93 (semantic).
\Cref{fig:likert} illustrates this shift clearly—baseline responses clustered around low-neutral ratings (2-4), while semantic system responses concentrated on high ratings (5-7).
The Transparent metric reinforced this finding ($p$ = .004, $r$ = 0.87; Semantic: \texttt{M} = 5.64, Baseline: \texttt{M} = 3.79), indicating that participants found the semantic representation made AI decision-making more comprehensible.
Collaborative scores ($p$ = .005, $r$ = 0.80) improved from \texttt{M} = 4.43 to \texttt{M} = 6.00, suggesting that participants perceived the system as shifting from an opaque generator into an understandable partner.

\paragraph{Improving Iterative Refinement and Collaborative Experience.}
Participants reported that the semantic structure made the refinement process more transparent and manageable.
They reported significantly higher Ease of Modification ($p$ = .010, $r$ = 0.71), with scores increasing from \texttt{M} = 3.71 (baseline) to \texttt{M} = 6.00 (semantic), indicating that they felt more capable of refining designs.
The response distribution in~\Cref{fig:likert} reveals that baseline users struggled with modifications, while semantic system users found refinement straightforward (predominantly 5-7 ratings).
Modification Reflection ($p$ = .032, $r$ = 0.61; Semantic: \texttt{M} = 5.57, Baseline: \texttt{M} = 4.21) showed the smallest but still significant effect size, suggesting that while semantic structure improved perceived refinement predictability, some challenges remain.
The Controllability metric ($p$ = .006, $r$ = 0.73; Semantic: \texttt{M} = 5.64, Baseline: \texttt{M} = 3.86) suggested that semantic-level manipulation provided a greater sense of agency over iterations.
Finally, Match Goal scores ($p$ = .045, $r$ = 0.59; Semantic: \texttt{M} = 5.64, Baseline: \texttt{M} = 4.86) indicated that the cumulative effect of these improvements helped users feel they could better achieve their intended designs through the iterative process.

\subsubsection{Qualitative Results}

We conducted a hybrid thematic analysis (deductive RQs and inductive coding) on usage patterns and interview responses.
Themes are organized by the comparative lens: expressivity/entry, interpretability/feedback, iteration/control, cognitive load/learnability, adoption/workflow, and challenges.

\paragraph{Articulating Intent (RQ1): structure turns vague ideas into editable levers.}
Participants consistently reported that semantic prompting—with auto-extracted fields, value recommendations, and slot-level editing—helped them say what they meant.
Several described the experience as being ``told what to specify'' and ``realizing what I forgot to mention.''
The system’s semantic attributes (\eg, typography, interaction, information architecture elements) helped participants who struggled with prompt formulation: ``I could just accept good suggestions and move on'' [P5]; ``This gives me the hints for what I want'' [P13].
Also, participants appreciated the automatic mapping into semantic slots: ``My paragraph became checkable items I can tweak.''

Regarding parsing accuracy, twelve participants generally found the natural language to semantic conversion accurate and helpful.
P5 characterized the parsing function as ``organizing my thinking for me,'' while P11 was ``surprised by how appropriate the categorization was.''
Eleven participants also appreciated how the system flagged empty fields in relationship analysis (Needs Value), guiding them to consider overlooked or missing details.
P4 described how this feature guided decision-making: ``I knew I wanted a brighter design but wasn't sure which exact one. The relation analysis recommended colors based on my `Target User' and `Goal,' so I applied it immediately.''
P7 noted that ``empty fields and suggested value showed me what was missing from my prompt,'' enabling iterative completion.

\paragraph{Understanding outputs (RQ2): traceability improves evaluation and trust.}
The semantic extraction and relationship analysis made system behavior legible, with participants reporting high perceived accuracy in the system's analysis of generated UI.
Participants valued analyzed semantics and match/conflict/omission checks as honest coaches that clarified both why the UI looked a certain way and what to do next [P1, P3, P4, P5, P10, P12].

Regarding semantic-to-UI adherence, participants reported that specified values were generally well-reflected.
P2 noted that ``since the system tells me what values to input, I could specify clearly, and it felt well-reflected,'' while P10 observed that abstract specifications like ``warm pastel tones'' resulted in coherent choices across related elements and generated UI.
However, adherence varied with input specificity—``clearly specified requests came out as expected, while vague requests produced vague results'' [P14]—and design novelty, with familiar UI patterns yielding higher fidelity than novel concepts [P6].
The augmented semantics (analyzing implementation) received positive feedback for extraction accuracy.
P8 reported that ``the analyzed results were quite accurate'' and P4 noted ``interaction details I would otherwise forget,'' and P5 found augmented semantics helpful for discovering implicit decisions, making it easy to map design intent to which part of the UI.
For semantic relationship analysis, conflict and match value indicators were perceived as actionable and reliable in understanding generated UI.
P11 remarked that the ``red conflict lines'' acted as an immediate feedback loop, ``telling me what to fix first'' without needing to guess why the UI looked wrong. P10 added that seeing the connections helped explain design choices.
Card-level feedback and one-click acceptance/replacement were frequently praised for reducing evaluation effort: ``I can see which intent maps to which part of the screen'' [P3].
Several participants linked this transparency to greater trust and faster decisions.

\paragraph{Iterative refinement (RQ3): scoped edits stay put; drift is contained.}
The most distinct advantage of the semantic system was its ability to support iterative refinement, reducing ``semantic drift''—the tendency for AI models to unexpectedly alter unrelated elements during updates.
In the baseline chat condition, five participants (P4, P8, P11, P13, P14) explicitly described ``ripple effect'' failures.
P14 noted: ``I asked to change the top section... but it randomly changed the button colors at the bottom too.''
P8 echoed this, stating that ``Chat updates often reset previous design choices.''
Nine participants highlighted the semantic structure's ability to enable ``scoped edits,'' allowing users to lock in specific decisions.
P2 and P9 highlighted the accuracy of these targeted updates, noting the ability to ``target just the node I want to change,'' ensuring that the rest of the interface remained stable.
The accept/replace workflow of augmented semantics and relationships was described as ``clickable evolution'' that sped up iteration [P5, P8].
In contrast, several noted that chat-based refinement sometimes drifted, altering unrelated visual attributes or layout despite narrow requests [P4, P8, P11, P13, P14].
Participants requested lightweight change auditing—``show me exactly what changed and why''—and simple rollback/version snapshots [P1, P8, P12]. They framed these as additive features atop an already predictable loop.

\paragraph{Cognitive load and learnability: brief onboarding, sustained payoff.}
After a short acclimation, participants described the semantic surface as lowering effort via problem ``chunking'' (content, function, components, style) and next-step signaling (warnings and recommendations) [P4, P5, P8, P11]. 
Also, P8 credited the system with saving time by surfacing interaction details they would otherwise forget. However, information density (multiple analysis entry points, new terms) led to a brief learning curve for some [P5, P11, P14].
Participants suggested progressive disclosure of semantics and inline examples/tooltips to smooth first use.

\paragraph{Adoption and workflow fit: suitable for ideation-to-refinement and mixed-skill teams.}
Across roles, participants saw the semantic approach as especially well-suited to early product exploration, structured refinement, and cross-functional collaboration.
Non-designers highlighted gains in expressing and checking intent; designers and engineers pointed to component-level semantics as helpful for handoff and consistency [P9, P11, P12, P13].
Several participants wanted a bridge to lightweight direct manipulation of elements in generated UI that writes back to semantics [P4, P7, P8, P10].

\paragraph{Challenges and Error Responses.}
Participants identified several limitations in our system. Initial confusion around semantic category boundaries was indicated—P2, P6, and P7 reported difficulty distinguishing Design Style from Visual Mood, or Function from Information Architecture. These concerns typically subsided after one or two generation cycles as participants developed intuitions for the categories. For the semantic parsing, some participants noted that long natural-language descriptions sometimes ``thinned out'' when compressed into semantic slots, suggesting nuance loss in parsing. Regarding relationship analysis, P1 expressed skepticism about trustworthiness (``Where's the evidence or source?''), while P11 observed occasional cases where the system ``focuses on specific intent, not for the whole intent.'' Several participants [P3, P5, P7, P11] noted that both systems excel at conventional, minimal styles but struggle with highly branded or stylistically distinctive designs. These findings indicate areas for improvement in semantic granularity, parser fidelity, and support for creative divergence.

%% file: section/7-discussion.tex
\section{Discussion}

\subsection{Semantics as Intermediate Representation}
Our study examined how our semantic layer mediates between natural-language intents and UI artifacts across three recurring challenges: articulating intent, interpreting model outputs, and iteratively refining designs.
First, explicit semantic slots and relations scaffold intent expression, reducing under-specification that commonly leads to prompt thrashing. Second, bidirectional mappings (intent~$\leftrightarrow$~artifact) make generated outputs interpretable by allowing back-tracing from components to the meanings that shaped them. Third, scoped editing localizes the impact of changes (local/section/global), stabilizing iterations and mitigating semantic drift. Taken together, these results suggest that semantic mediation can bridge both the \emph{gulf of execution} (what to say to a model) and the \emph{gulf of evaluation} (how to read what it produced) by inserting a usable intermediate representation between people and generative systems.

\subsubsection{Why Semantic Mediation Works}
We propose three complementary mechanisms. First, \textit{Structure-preserving input.} Typed slots and explicit relations capture constraints that linear prompts often compress or omit, improving the fidelity of model conditioning. Second, \textit{Traceability through bidirectionality and relationships.} Aligning artifacts to semantics enables inspection, diagnosis, and principled revision: when a card’s visual or behavioral property is off-spec, the corresponding semantic node is immediately discoverable. Third, \textit{Locality with scoped edits.} Designers can target changes to a bounded region by changing the corresponding semantics, which reduces collateral changes elsewhere and supports reasoned ``what-if'' exploration. We also observed boundary cases that illuminate these mechanisms: when tasks demand highly bespoke visual styles or idiosyncratic brand moves, the semantic layer may bias toward conservative convergence; in such cases, lightweight override channels and style exemplars remain important.

\subsubsection{Scope and Generalizability}
The approach is most effective for interface families with stable components, reusable patterns, and articulated design-system constraints (\eg, dashboards, form-intensive workflows, onboarding flows). It generalizes to adjacent activities where semantics are likewise structured, such as layout variation within a design system and accessibility conformance checks (\eg, contrast, focus states). Organizational adoption will be smoother when the semantic layer is integrated with existing design-system tokens and version-control practices, enabling provenance, review, and rollback at the level of meanings rather than only pixels or code diffs.

\subsection{Design Implications}
Grounded in our findings, we consolidate the design space into four implications that translate semantic mediation into practice. Each implication targets a specific gulf (execution or evaluation) or the amplification problem, and pairs explanation with guidance.

\paragraph{Hybrid Entry for Exploration-to-Convergence.}
Participants valued the low-friction start of chat yet relied on semantic controls for precision. A productive flow keeps free-form prompts for ideation and \emph{auto-maps them to editable semantics} for convergence. After a prompt, surface an \emph{Intent Draft} that proposes high-confidence fields (\eg, target user, task, components, style) with inline acceptance toggles and quick-pick recommendations. Keep the raw prompt visible (hover/expand) to avoid perceived context loss. This arrangement lets people ``say it like a person,'' then ``tune it like a designer,'' reducing formulation effort while preserving control. Begin with a compact slot set and expand coverage as parsers mature.

\paragraph{Actionable Transparency over Explanations.}
Explanations were most helpful when they immediately suggested the next edit. Instead of static rationales, present \emph{card-level feedback} bound to specific semantic nodes—Matches (collapsible evidence), Conflicts (with one-click fixes), and Omissions (with autosuggested values). Clicking ``Fix'' should commit a scoped change and annotate a change log. Brief, situated microcopy (\eg, ``Color mood conflicts with target age; try ‘muted pastel’?'') clarifies causality while shortening the path from insight to action. Ship cards-first; reserve graph views as an advanced toggle that links back to the same fixes.

\paragraph{Locality by Default, with Narrated Change.}
One of the most celebrated gains was predictable locality: edits should stay where users intend. Make \emph{local scope} the default and require escalation to section or global. An \emph{Edit Scope} switch (Local \(\triangleright\) Section \(\triangleright\) Global) on every control, paired with an \emph{impact preview} (affected nodes count), could keep changes bounded. Each commit emits a diff chip (\eg, ``Button corner radius 8~$\rightarrow$~4'') and updates a reversible change log. This narrative scaffolds trust, addresses ``what changed and why,'' and directly tackles amplification by containing side effects.

\paragraph{A Bridge to Direct Manipulation that Writes Back to Semantics.}
Designers often want to nudge spacing, alignment, or emphasis directly on canvas without losing traceability. Provide lightweight direct manipulation—handles for spacing/alignment, sliders for radius/elevation, swatch pickers—that \emph{update the bound slot} and record a semantic diff. Offer ``Lock to Design System'' so tweaks snap to tokenized ranges. This bridge satisfies craft instincts while maintaining explainability and repeatability; start with a small set of high-impact properties and expand after stability testing.

\subsection{Limitations and Future Work}
Our investigation has several limitations.
First, the semantic vocabulary introduces an initial learning curve; newcomers must internalize terms and relations.
Progressive disclosure, examples-as-templates, and automatic slot suggestions may alleviate this burden.
Second, slotting long-form intents risks nuance compression.
Dual views that co-display the original text and its structured interpretation, along with preservable ``verbatim'' spans for critical phrasing, can mitigate loss.
Third, the breadth of the evaluation remains limited: our evaluation mainly focused on perceived control, reflecting our theoretical focus on the \textit{gulfs} as cognitive gaps in the user's mental model.
Consequently, the artifact-based differences, as well as longitudinal and team-level cooperative conditions, remain to be empirically validated.
Additionally, technical constraints shaped the current implementation.
Our system covers a single React component, which helps reduce integration complexity but limits applicability to complex UI architectures.
Also, hallucination (including syntax errors in UI code) and unreliable localized editing remain challenges: instead of modifying only the intended region, the model may regenerate or alter unaffected content. Semantic diffs helped constrain these behaviors, but they served as a partial mitigation rather than a robust editing mechanism.

Future work follows tracks corresponding to these limitations.
First, longitudinal and organizational studies should examine how the semantic-based approach is utilized in actual design workflows involving complex, multi-screen products, including collaboration among designers, PMs, and engineers.
Second, advances in compliance by construction can integrate global tokens and accessibility rules so that violations are flagged and fixed directly in the semantic layer.
In parallel, technical extensions are needed to address scalability and model reliability.
Supporting multi-component, large-scale architectures will be essential for enabling real-world, long-term use in large and evolving UI codebases.
Closed-loop learning would treat accepted edits as training signals, allowing recommender models to improve fidelity and reduce correction effort over time.
Finally, cross-modal semantics could extend beyond text to semantics inferred from sketches, wireframes, or component trees, widening on-ramps into the structured layer while maintaining traceability.

%% file: section/8-conclusion.tex
\section{Conclusion}

In this paper, we explored how exposing semantic representations as an intermediate layer between natural language prompts and AI-generated UIs can help bridge the gulfs in generative UI design.
By conducting a thematic analysis of prompting guidelines, we identified a four-level hierarchical framework (Product, Design System, Feature, Component) that structures the complex, interdependent information required for UI generation.
Our system operationalizes this framework through three mechanisms: structured specification that guides users in articulating design intent, transparent analysis that reveals how AI interprets and implements semantics, and relationship-aware refinement that maintains consistency across iterations.
The comparative user study with 14 practitioners showed significant improvements across all dimensions—intent expression, output interpretation, and iterative refinement—with participants reporting more predictable and controllable design outcomes compared to traditional chat-based interfaces.
These findings suggest that making the implicit semantic layer explicit supports a shift in UI generation from an opaque, trial-and-error process into a systematic, interpretable workflow where users can effectively communicate with and understand AI systems.
While the approach introduces some initial learning overhead and may constrain highly creative explorations, it provides a practical path forward for democratizing UI design through AI while maintaining the precision and control that professional design work demands.